\documentclass[fleqn,usenatbib]{mnras}

\usepackage{newtxtext,newtxmath}

\usepackage[T1]{fontenc}

\DeclareRobustCommand{\VAN}[3]{#2}
\let\VANthebibliography\thebibliography
\def\thebibliography{\DeclareRobustCommand{\VAN}[3]{##3}\VANthebibliography}

\usepackage{graphicx}	
\usepackage{amsmath}	
\usepackage{amssymb}	
\usepackage{subcaption}
\usepackage{xcolor}
\definecolor{ao}{rgb}{0.0, 0.5, 0.0}
\definecolor{bv}{rgb}{0.54, 0.17, 0.89}
\definecolor{r}{rgb}{0.8, 0.0, 0.0}
\definecolor{notegreen}{rgb}{0.235,0.651,0.282}

\def\hi{\mbox{\,H\,{\sc i}}}

\def\lyafesc{$f_{\rm{esc}}^{\rm{Ly\alpha}}$}
\def\lycfesc{$f_{\rm{esc}}^{\rm{LyC}}$}

\newcommand{\halpha}{\mbox{\,H\,{\sc $\alpha$}}}
\newcommand{\lya}{\mbox{\,Ly\,{\sc $\alpha$}}}

\def\wlya{$W_{\lambda}$(Ly\,$\alpha$)}
\def\wha{$W_{\lambda}$(H\,$\alpha$)}

\title[The Ly$\alpha$-LyC connection at $z\simeq 4-5$]{Connecting the escape fraction of Lyman-alpha and Lyman-continuum photons in star-forming galaxies at $\mathbf{z\simeq 4-5}$}
\author[R. Begley et al.]{R. Begley$^{1}$\thanks{E-mail:rbeg@roe.ac.uk}, F. Cullen$^{1}$, R. J. McLure$^{1}$, A. E. Shapley$^{2}$, J. S. Dunlop$^{1}$, A. C. Carnall$^{1}$, D. J. McLeod$^{1}$, \and C. T. Donnan$^{1}$, M. L. Hamadouche$^{1}$ and T. M. Stanton$^{1}$\\\\
$^{1}$Institute for Astronomy, University of Edinburgh, Royal Observatory, Edinburgh EH9 3HJ, UK\\
$^{2}$Department of Physics \& Astronomy, University of California, Los Angeles, 430 Portola Plaza, Los Angeles, CA 90095, USA
}

\date{Accepted 2023 October 30. Received 2023 October 09; in original form 2023 June 05}

\pubyear{2023}

\begin{document}
\label{firstpage}
\pagerange{\pageref{firstpage}--\pageref{lastpage}}
\maketitle

\begin{abstract}
We present a study of the connection between the escape fraction of Lyman-alpha ($\mathrm{Ly\,\alpha}$) and Lyman-continuum  (LyC) photons within a sample of $N=152$ star-forming galaxies selected from the VANDELS survey at $3.85\leq z_{\rm{spec}}\leq 4.95$ ($\langle z_{\mathrm{spec}}\rangle=4.36$).
By combining measurements of H$\,\alpha$ equivalent width $(W_{\rm{\lambda}}(\rm{H\,\alpha}))$ derived from broad-band photometry with measurements of Ly$\,\alpha$ equivalent width $(W_{\rm{\lambda}}(\rm{Ly\,\alpha}))$ from the VANDELS spectra, we individually estimate $f_{\rm{esc}}^{\rm{Ly\alpha}}$ for our full sample. In agreement with previous studies, we find a positive correlation between $W_{\rm{\lambda}}(\rm{Ly\,\alpha})$ and $f_{\rm{esc}}^{\rm{Ly\alpha}}$, with $f_{\rm{esc}}^{\rm{Ly\alpha}}$ increasing from $f_{\rm{esc}}^{\rm{Ly\alpha}}\simeq0.04$ at $W_{\rm{\lambda}}(\rm{Ly\,\alpha})=10$ \AA\ to $f_{\rm{esc}}^{\rm{Ly\alpha}}\simeq0.1$ at $W_{\rm{\lambda}}(\rm{Ly\,\alpha})=25$ \AA. For the first time at $z\simeq4-5$, we investigate the relationship between $f_{\rm{esc}}^{\rm{Ly\alpha}}$ and $f_{\rm{esc}}^{\rm{LyC}}$ using $f_{\rm{esc}}^{\rm{LyC}}$ estimates derived using the equivalent widths 
of low-ionization, far-UV absorption lines in composite VANDELS spectra. Our results indicate that $f_{\rm{esc}}^{\rm{LyC}}$ rises monotonically with $f_{\rm{esc}}^{\rm{Ly\alpha}}$, following a relation of the form \lycfesc$\simeq 0.15^{+0.06}_{-0.04}$\lyafesc. Based on composite spectra of sub-samples with approximately constant \wlya, but widely different \lyafesc, we demonstrate that the \lycfesc$-$\lyafesc correlation is not driven by a secondary correlation between \lyafesc\, and \wlya. The observed \lycfesc$-$\lyafesc\, correlation is in good qualitative agreement with theoretical predictions and provides further evidence that estimates of $f_{\rm{esc}}^{\rm{LyC}}$ within the Epoch of Reionization should be based on proxies sensitive to neutral gas density/geometry and dust attenuation.

\end{abstract}

\begin{keywords}
galaxies: high-redshift -- galaxies: fundamental parameters -- intergalactic medium
\end{keywords}



\section{Introduction}\label{sec:intro}

The formation of the first stars and the growth of galaxy progenitors in the early Universe signalled the beginning of the Epoch of reionization (EOR), during which the fully neutral Hydrogen gas in the intergalactic medium (IGM) became completely ionized \citep[e.g.,][]{robertson+15}. Although it is generally acknowledged that reionization was completed between $z\sim5-6$, based primarily on measurements of the Ly$\alpha$ forest from distant quasars \citep[e.g.,][]{fan+06,mcgreer+15,goto+21},
key details of the process of reionization and the nature of the sources responsible remain a matter of debate \citep[e.g.,][]{mason+19,naidu+20}.

One fundamental reason for the continued debate is that the progress of reionization is intimately linked to the physical properties of the sources that are responsible for it, as well as their location within the large-scale structure of the Universe \citep{robertson+21}. It is now widely accepted that active galactic nuclei are simply too rare at high redshift to contribute significantly to reionization \citep[e.g.,][]{matsuoka+23} and that the dominant contribution to the ionizing photon budget must come from star-forming galaxies (SFGs) \citep[e.g,][]{robertson+15,chary+16,iwata+22}.

A key component needed to quantify the ionizing photon budget is the abundance of star-forming galaxies during the EOR, most commonly parameterised through the UV luminosity density ($\rho_{\mathrm{UV}}$), that is now being established with increasing accuracy out to $z\simeq12$ by JWST \citep[e.g.,][]{donnan+22,finkelstein+22b,harikane+22}. The latest JWST results indicate that $\rho_{\mathrm{UV}}$ displays a smooth, steady decline through the EOR \citep[e.g.,][]{donnan+23, mcLeod+23} indicating, in principle, the availability of more ionizing photons than some pre-JWST studies \citep[e.g.,][]{Oesch+18, Ishigaki+18} had suggested. 

Another essential element is the ionizing photon production efficiency $\xi_{\mathrm{ion}}$, a measure of the number of ionizing photons produced per unit UV luminosity of the star-forming galaxy population. Recent evidence has suggested that the faint, blue population of galaxies commonly found at $z>6$ \citep[i.e. with UV spectral slope  $\beta\lesssim-2.3$;][]{cullen+22} display $\mathrm{log}(\xi_{\rm{ion}})$ in the range $\sim25.5-25.8$ \citep{stark+15,bouwens+16}, a factor of $\geq 2-3$ higher than canonically assumed in models of reionization \citep[e.g.,][]{bouwens+12b,finekelstein+12b,duncan+15}. 

Although the abundance of SFGs and the elevated values of $\xi_{\mathrm{ion}}$ at $z>6$ indicate that sufficient numbers of ionizing photons are being generated within the EOR, whether or not these Lyman-continuum (LyC) photons escape their source galaxies and ionize the IGM is ultimately determined by the escape fraction ($f_{\rm{esc}}^{\rm{LyC}}$).\\

Due to the near-total attenuation of UV photons below the Lyman break by the IGM at $z\geq 4$ \citep[e.g., ][]{steidel+18}, direct observational constraints on $f_{\rm{esc}}^{\rm{LyC}}$ are only possible up to $z\simeq 3.8$. At the highest redshifts where such observations are possible, recent estimates from deep photometric and spectroscopic studies have typically constrained the average $f_{\rm{esc}}^{\rm{LyC}}$ to be $\simeq 5-10$\% \citep[e.g.,][]{steidel+18,pahl+21,mestric+21,begley+22,saldana-lopez+22b}. Generally, these studies have also found that the fainter, less-dust-obscured galaxies that are expected to be numerous during the EOR are more likely to display high $f_{\rm{esc}}^{\rm{LyC}}$. 
Promisingly, these trends are in accordance with the assumptions that are often made in reionization models \citep{robertson+10,wise+14,robertson+15,finkelstein+19} where $\langle f_{\rm{esc}}^{\rm{LyC}} \rangle \geq 5$\% is typically required for reionization to be completed by $z\sim5-6$.

In spite of these encouraging results, we still lack a comprehensive understanding of exactly {\it how} LyC photons escape galaxies. To this end, a number of studies have attempted to link $f_{\rm{esc}}^{\rm{LyC}}$ to nebular emission-line features, such as Mg{\mbox{\,\sc{ii}}} \citep{katz+22}, C{\mbox{\,\sc{iv}}} \citep{schaerer+22} and [O{\mbox{\,\sc{iii}}}] \citep[e.g.,][]{verhamme+15,wang+19,izotov+20,nakajima+20}, which are modulated by the same interstellar medium (ISM) and stellar-population properties that ultimately determine $f_{\rm{esc}}^{\rm{LyC}}$.

One of the most promising and closely investigated indicators is the {\lya} emission line, with a number of studies showing that strong {\lya} emission (i.e., identified by a high {\lya} equivalent width, \wlya) is an excellent indicator of high $f_{\rm{esc}}^{\rm{LyC}}$ \citep{marchi+18,pahl+21,begley+22}. 
Furthermore, a clear correlation between $f_{\rm{esc}}^{\rm{Ly\alpha}}$ and $f_{\rm{esc}}^{\rm{LyC}}$ has been found by \cite{flury+22b}, using galaxies at $z\simeq0.3$ selected from the Low Redshift Lyman Continuum Survey (LzLCS;  \citealt{flury+22a}); a finding consistent with simulation results \citep[e.g,][]{dijkstra+16,kimm+19}. 

This \lya$-$LyC connection is expected given that $f_{\rm{esc}}^{\rm{Ly\alpha}}$ and $f_{\rm{esc}}^{\rm{LyC}}$ are both modulated by the geometry and nature of the gas in the vicinity of young star-forming regions \citep[e.g.,][]{shapley+03,chisholm+18,gazagnes+20,maji+22}. Furthermore, the characteristics of the young stellar populations themselves influence both LyC and {\lya} photon production, with low-metallicity stellar populations having higher $\xi_{\rm{ion}}$ and therefore producing more LyC and {\lya} photons for a given star-formation rate \citep[e.g.,][]{trainor+15,erb+16,trainor+16,cullen+20}.   

The main escape path for ionizing photons is likely through channels of low-column-density, and/or high-ionization-state gas in the ISM, through which {\lya} photons can also escape in significant quantities \citep[e.g.,][]{atek+08,dijkstra+16,jaskot+19}. \citet{gazagnes+20} presented strong observational evidence for LyC leakage via this mechanism, finding significant correlations between the presence of low H{\mbox{\,\sc{i}}} covering fractions and the observed LyC escape fraction. Similar correlations have been found in several other independent observational studies \citep[e.g.,][]{verhamme+17, chisholm+18, saldana-lopez+22a}.

This argument is further bolstered by the well-established links between {\lya} and the ISM properties of galaxies. \citet{shapley+03} showed that the observed range of {\wlya} in Lyman break galaxies at $z\sim3$ is accounted for by variations in the covering fraction of neutral outflowing H{\mbox{\,\sc{i}}} gas and dust \cite[see also;][]{atek+08,kornei+10,berry+12,reddy+16}. More-recent literature studies have uncovered a relationship between {\wlya} and the covering fraction of neutral gas, as traced by the strength of low-ionization-state ISM absorption lines \citep{henry+15,du+18,steidel+18,jaskot+19,trainor+19}. 

In this study, we explore the connection between {\lya} and LyC escape in galaxies only $\simeq 300$ Myr after reionization was completed, using a sample of galaxies in the range $3.85\leq z_{\rm{spec}}\leq4.95$ drawn from the VANDELS ESO public spectroscopic survey. We combine direct {\lya} line measurements from the VANDELS spectra with {\halpha} luminosity constraints based on robust SED fitting, allowing us to individually estimate $f_{\rm{esc}}^{\rm{Ly\alpha}}$ for each galaxy. Using composite spectra formed from sub-samples of {\lya} emitters, we then investigate how $f_{\rm{esc}}^{\rm{LyC}}$, as estimated from FUV LIS ISM line strengths, varies with $f_{\rm{esc}}^{\rm{Ly\alpha}}$ and {\wlya}.

The structure of the paper is as follows. In Section \ref{sec:data} we describe the VANDELS spectral dataset, the associated photometric catalogues and our sample selection criteria. In Section \ref{sec:lyafesc} we describe the methodology used to derive the individual $f_{\rm{esc}}^{\rm{Ly\alpha}}$ measurements. In Section \ref{results} we explore the $f_{\rm{esc}}^{\rm{Ly\alpha}}-$\wlya\, correlation, before presenting our constraints on $\langle f_{\rm{esc}}^{\rm{Ly\alpha}}\rangle$ and proceeding to constrain $f_{\rm{esc}}^{\rm{LyC}}$ from the strength of low-ionization-state absorption lines in composite VANDELS spectra.
We provide a discussion of our main results in Section \ref{sec:discussion} and present our conclusions in Section \ref{sec:summary}. Throughout the paper we adopt the following cosmological parameters: $H_{\rm{0}}=70\,\rm{km\,s^{-1}Mpc^{-1}}$, $\Omega_{\rm{m}}=0.3$, $\Omega_{\rm{\Lambda}}=0.7$ and all magnitudes are quoted in the AB system \citep{oke_gunn+83}. 
\section{Data and Sample Selection}\label{sec:data}
Our sample of star-forming galaxies is drawn from the VANDELS ESO public spectroscopic survey final data release (DR4, \citealt{garilli+21}). 
The VANDELS survey used the VIMOS spectrograph \citep{le-fevre+03} installed on the ESO {\it Very Large Telescope} (VLT) to obtain ultra-deep spectra of 2087 galaxies at red-optical wavelengths ($4800$ \AA $\ <\lambda<10000$ \AA) in the Chandra Deep Field South (CDFS) and UKIDSS Ultra Deep Survey (UDS) fields \citep{mclure+18, pentericci+18}. The vast majority of VANDELS targets were main-sequence star-forming galaxies at $2.4\leq z \leq 7.0$, for which the ultra-deep ($20-80$ hour integration) VIMOS spectra\footnote{The VANDELS observations used the MR grism$+$GG475 order sorting filter with 1 arcsec slit widths. Approximately $90\%$ of the observations had seeing of $\:\lesssim1$ arcsec.} cover the rest-frame far-ultraviolet (FUV), enabling measurements of the {\lya} line \citep[e.g.,][]{cullen+20}.  Full details of the survey design and target selection can be found in \citet{mclure+18}.

All galaxies targeted in VANDELS benefit from deep, multi-wavelength imaging data covering the observed wavelength range $0.38~\mu \mathrm{m} \lesssim \lambda_{\rm{obs}}~\lesssim~4.5~\mu \mathrm{m}$. Approximately half of the VANDELS sample lies within the CANDELS GOODS-S and UDS \emph{Hubble Space Telescope} (HST) imaging footprint \citep{koekemoer+11, grogin+11}, for which we adopt the photometry from \cite{guo+13} and \cite{galametz+13}, respectively. The other half of the sample lies outside the CANDELS footprint, but benefits from wider-area, primarily ground-based optical/nearIR imaging. For this study we adopt the updated VANDELS photometry catalogues described in \citet{garilli+21} and publicly released as part of DR4. Crucially, in addition to optical/near-IR data, the full VANDELS sample benefits from deep, deconfused {\it Spitzer} IRAC photometry  at 3.6 $\mu \mathrm{m}$ and 4.5~$\mu \mathrm{m}$. As discussed below, it is the photometric excess at 3.6 $\mu \mathrm{m}$ that provides our measurement of {\halpha} line flux, with the 4.5 $\mu \mathrm{m}$ photometry providing a long-wavelength (emission-line free) anchor for the SED fitting.

Our initial sample consists of all VANDELS galaxies in the redshift range $3.85 \leq z_{\rm spec} \leq 4.95$, within which the IRAC 3.6 $\mu \mathrm{m}$ filter is contaminated by the \halpha\, emission line. In addition, to ensure the redshifts are robust, we restricted the sample to the $N=263$ galaxies in this redshift range with redshift quality flags of $z_{\rm{flag}}=3,4 \ \mathrm{or} \ 9$, corresponding to a $\geq 95\%$ probability of being correct \citep{garilli+21}.
Finally, in order to measure \lya \ escape fractions, we restricted the sample to galaxies which displayed \lya \ in emission (i.e., an equivalent width $>0$ {\AA}; see below), resulting in a sample of $N=152$ galaxies with a median redshift of $\langle z\rangle=4.36$.

\begin{figure}
        \centerline{\includegraphics[width=\columnwidth]{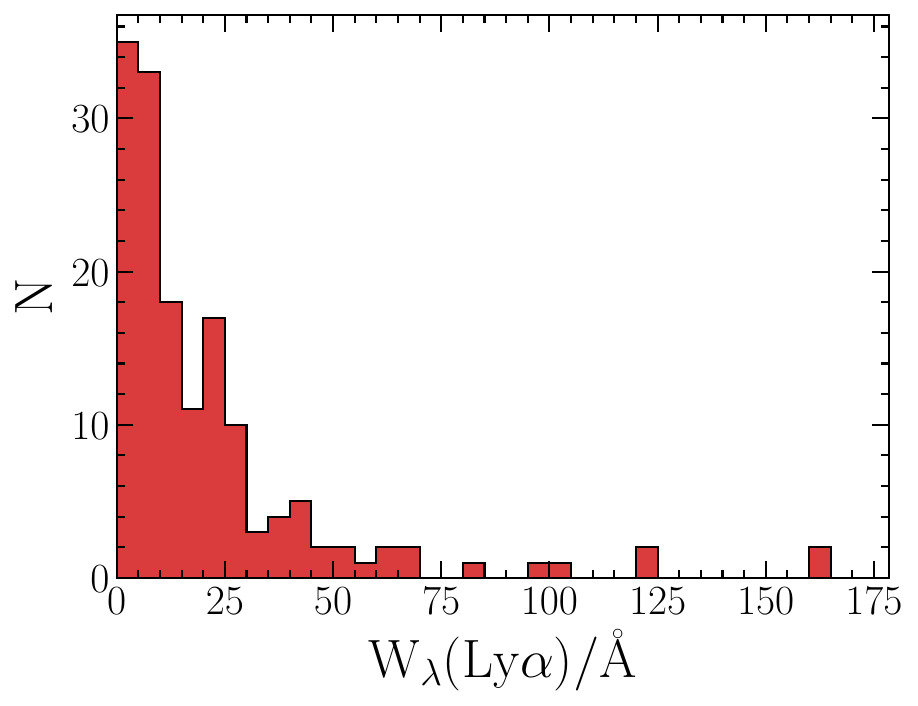}}
        \caption{The distribution of rest-frame {\lya} equivalent width for our final sample of $N=152$ star-forming galaxies at $3.85\leq z_{\rm{spec}}\leq 4.95$ showing \lya\, in emission. The median rest-frame equivalent width 
        is $\langle W_{\lambda}(\rm{Ly\,\alpha})\rangle~=~11.9$\,\AA\,.}
        \label{fig:sample_lyaew_hist}
\end{figure}

\section{Emission line flux measurements}\label{sec:lyafesc}
Our principal aim is to determine \lya\ escape fractions for the galaxies in our sample.
To do this we combine measurements of the \emph{observed} \lya\ flux (measured directly from the VANDELS spectra) with estimates of the \emph{intrinsic} flux derived from the observed \halpha \ flux (measured using the IRAC 3.6 $\mu\mathrm{m}$ flux excess). In this section, we describe each stage in the process of deriving our \lya\ escape fraction measurements for our sample of $3.85\leq z_{\rm spec}\leq 4.95$ galaxies.

\subsection{Observed \textbf{\lya} fluxes and equivalent widths}\label{subsec:observed_lya}
Observed \lya \ fluxes and rest-frame equivalent widths, \wlya, are measured from the VANDELS spectra following the method described by \citet{kornei+10} and adopted in our previous analysis of the correlation between \wlya\ and stellar metallicity \citep{cullen+20}. Briefly, the line flux is measured by integrating the spectrum 
around the peak of the \lya \ emission line, between limits defined as the wavelengths where the spectrum intersects the `red' and `blue' continuum levels, defined as the median flux between $1120$\AA$\, \leq\lambda_{\rm{rest}}\leq1180$\AA\ ($c_{\rm{blue}}$) and $1228$\AA$ \, \leq\lambda_{\rm{rest}}\leq1255$\AA\ ($c_{\rm{red}}$), respectively\footnote{For three objects no continuum was detected in the spectra and continuum levels were estimated from the best-fitting SED model (Section \ref{subsubsec:sed_fitting}).}. The rest-frame equivalent width is then simply obtained by dividing the integrated Ly$\alpha$ line flux by $c_{\rm{red}}(1+z)$. 

For each galaxy, the above process is repeated $500$ times, each time perturbing the galaxy spectrum on a pixel-by-pixel basis by its corresponding error value. From the resulting distribution, the median and scaled median absolute deviation ($\sigma_{\rm{MAD}}\simeq1.4826\times\rm{MAD}$) are calculated and adopted as the final rest-frame \wlya\, and uncertainty. The equivalent width distribution of our final sample of $N=152$ galaxies showing \lya\, emission is shown in Fig. \ref{fig:sample_lyaew_hist}.

\subsection{Observed H$\alpha$ fluxes and equivalent widths}

\begin{figure}
    \centerline{\includegraphics[width=\columnwidth]{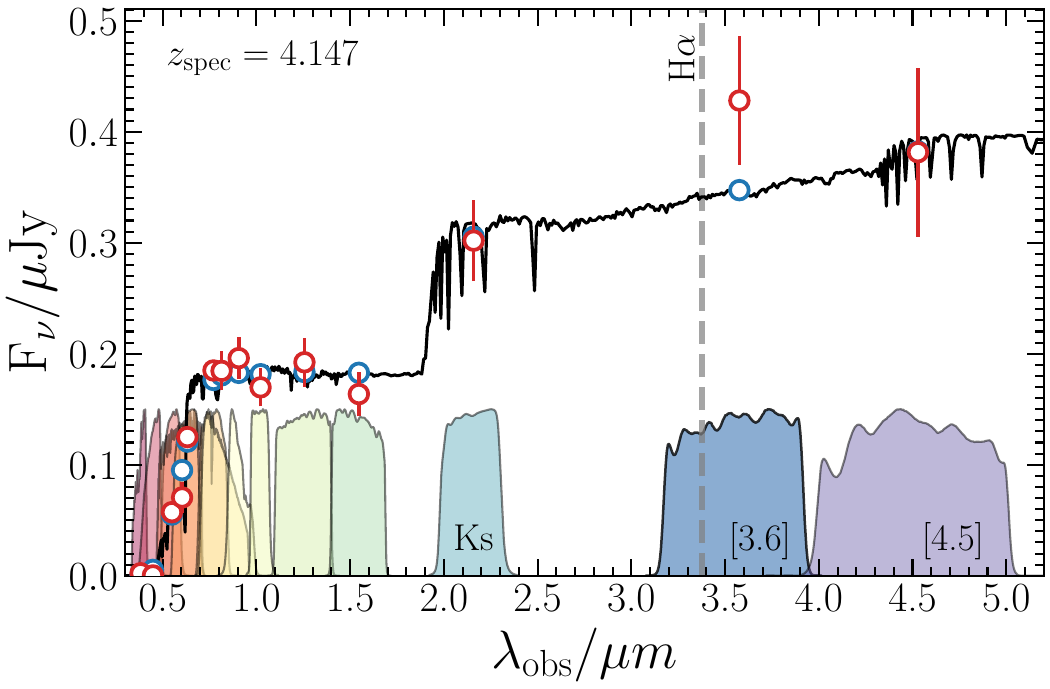}}
    \caption{The best-fitting SED model from \textsc{fast++} for an example galaxy at $z_{\rm{spec}}~=~4.147$. 
    SED fits to the observed fluxes (red points) excluding the IRAC 3.6 $\mu\mathrm{m}$ filter which is contaminated by {\halpha} emission (dashed grey line) allows the excess between the observed and predicted 3.6 $\mu\mathrm{m}$ flux (blue points) to be measured. In this example,
    the flux excess is $\Delta\rm{mag}=-0.22^{+0.16}_{-0.13}$, corresponding to \wha$=250\pm170${\,\AA}. 
    The normalised transmission profiles for the multi-wavelength photometry available for this galaxy are shown in the lower half of the panel.}
    \label{fig:ha_sed_demo}
\end{figure}

To estimate the \emph{intrinsic} \lya \ flux we first obtain an estimate of the observed \halpha \ flux \ and nebular dust attenuation (see also; Section \ref{subsec:sfr_rates})
from SED fitting the available multi-wavelength photometry. When fitting the photometry we exclude the IRAC 3.6 $\mu\mathrm{m}$ filter containing the \halpha \ line, which enables a robust estimate of the \wha\ and \halpha\ line flux, via the well-known photometric excess technique \citep[Fig. \ref{fig:ha_sed_demo}; e.g.,][]{stark+13,bouwens+16,marmol-queralto+16,smit+16}. 
Below we give details of the stellar population modelling and our method for deriving \wha.

\subsubsection{Stellar population modelling}\label{subsubsec:sed_fitting}
In this analysis we use the \textsc{fast++} code \citep[][]{kriek+09,schreiber+18} to perform SED fitting for each galaxy, using 
\citet{bruzal-charlot+03} stellar population synthesis models with a \citet{chabrier+03} IMF and a metallicity range of 
$0.2-0.4\times\rm{Z}_{\odot}$. We assume a constant star-formation history, with the age allowed to vary within the range $6.7\leq \rm{log(t/yr)}\leq10.0$ in steps of $\Delta(\rm{log(t/yr)})=0.2$. We adopt the \citet{calzetti+00} dust attenuation law and allow 
the absolute attenuation $A_{\rm{V}}$ to vary within the range $0.0\leq A_{\rm{V}}\leq 4.0$. All photometric data points are included
in the SED fitting, except for the IRAC 3.6 $\mu\mathrm{m}$ filter, and we do not include nebular emission in the SED fits (see Fig. \ref{fig:ha_sed_demo} for an example). Our final sample has a median stellar mass of $\langle \rm{log(} M_{*}/\rm{M_{\odot}})\rangle=9.08$ and a median star-formation rate of $\langle \rm{log(} {\rm SFR}/\rm{M_{\odot}yr^{-1}})\rangle=1.12$, fully consistent with being located on the star-forming main sequence at $z\simeq4-5$.

\subsubsection{\halpha \ equivalent widths}\label{subsubsec:ha_constraints}
\begin{figure}
    \centerline{\includegraphics[width=1.05\columnwidth]{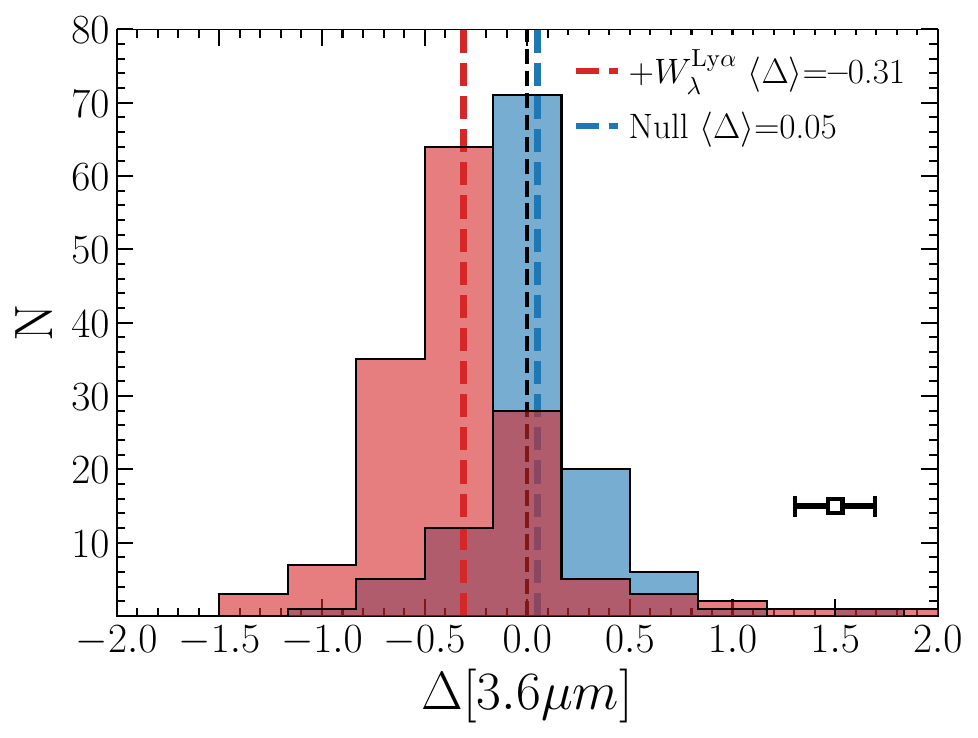}}
    \caption{The distribution of $\Delta[3.6\mu\rm{m}]$, defined as the difference between the observed magnitude in the 3.6 $\mu$m filter and the predicted magnitude from the best-fitting SED model (see text for details). The red histogram shows the distribution of the $N=152$ galaxies with \lya\ in emission. The median value of $\Delta[3.6\mu\rm{m}]=-0.31$, corresponds to $\langle W_{\lambda}(\rm{H\,\alpha})\rangle\simeq365$ {\AA} at the median redshift of the sample. The blue histogram shows the distribution for the null sample, for which the IRAC 3.6 $\mu$m filter is free from \halpha\ contamination. As expected, the null distribution is consistent with no flux excess. The $\Delta[3.6\mu\rm{m}]$ typical error is denoted by the black error bar.}
    \label{fig:Ha_hist}
\end{figure}

For each galaxy in our sample the \halpha \ equivalent width is estimated by comparing the observed IRAC 3.6 $\mu\mathrm{m}$ flux 
to the stellar population model prediction (see Fig. \ref{fig:ha_sed_demo}).
For each galaxy we determine
\begin{equation}
    \Delta[3.6\mu \mathrm{m}]=\rm{m_{obs}^{3.6\mu \mathrm{m}}}-\rm{m_{mod}^{3.6\mu \mathrm{m}}},
\end{equation}
where $\rm{m_{obs}^{3.6\mu \mathrm{m}}}$ and $\rm{m_{mod}^{3.6\mu \mathrm{m}}}$ are the observed and model apparent magnitudes in the IRAC 3.6$\mu\mathrm{m}$ filter, respectively.
The resulting distribution of $\Delta[3.6\mu \mathrm{m}]$ is shown in Fig. \ref{fig:Ha_hist}.
The distribution shows a clear systematic shift from $\Delta[3.6\mu \mathrm{m}]=0$ towards negative values (median $\Delta[3.6\mu \mathrm{m}] = -0.31$), signifying the presence of \halpha \ emission in the majority of our sample.  
Based on the IRAC 3.6 $\mu\mathrm{m}$ excess, we estimate the rest-frame \halpha \ equivalent width using
\begin{equation}
    W_{\lambda}({\rm{H\,\alpha}}) = f_{{\rm{H\alpha}}} \times \frac{ \mathcal{W}_{\rm{eff}}^{3.6\mu{m}} }{ 1+z_{\rm{spec}} } \times \left( 10^{-0.4\times\Delta[3.6\mu{m}]} -1 \right),
\end{equation}
where $\mathcal{W}_{\rm{eff}}^{3.6\mu \mathrm{m}}=6844$ \AA\, is the effective width of the 3.6 $\mu \mathrm{m}$ filter and $f_{\rm{H\,\alpha}}$ is the fraction of the total contaminating line flux attributed to the \halpha \ line alone (i.e. excluding [\mbox{\,N\,{\sc ii}}] $\lambda6584$ \AA\ and [\mbox{\,S\,{\sc ii}}] $\lambda\lambda6717,6731$ \AA). The value of $f_{\rm{H\,\alpha}}$ is often cited to be in the range $f_{\rm{H\,\alpha}} \simeq 0.71-0.9$ \citep[e.g.,][]{shim+11,stark+13,marmol-queralto+16, smit+16}. In this analysis, we use a fiducial conversion of factor of $f_{\rm{H\,\alpha}} = 0.84$, following \citet{smit+16}.

The median value of $\Delta[3.6\mu \mathrm{m}] = -0.31$ corresponds to $W_{\lambda}({\rm{H\,\alpha}})=365$ \AA\, at the median redshift of $z=4.36$. This value is in excellent agreement with previous estimates at similar redshifts.
For example, \citet{smit+16} find {\wha\,}$=325\pm22$\,\AA\, for a sample of $N=80$ spectroscopically confirmed galaxies at $3.8\leq z\leq5.0$. In the same redshift range, \citet{stark+13} report values of \wha\,$\simeq280-410${\,\AA} (assuming $f_{\rm{H\alpha}}=0.76$).
Finally, we note that our constraints are also in good agreement with the redshift evolution for {\wha} derived by \citet{marmol-queralto+16} across the redshift range $1\leq z\leq 5$.

\subsubsection{Null Sample Verification}\label{subsubsec:null_sample}
As a sanity check of our method, we perform the same analysis on a sample of $N=119$ VANDELS star-forming galaxies selected in the redshift range $3.6\leq z_{\rm{spec}}\leq 3.8$, within which the \halpha \ emission line does not contaminate the IRAC 3.6 $\mu$m photometry and a photometric excess signature should not be detected\footnote{We note that this redshift range has been chosen such that none of the photometric filters are affected by nebular emission line contamination.}. We apply the same procedures discussed above to this `null' sample, deriving the blue histogram in Fig. \ref{fig:Ha_hist}. The null sample distribution is fully consistent 
with $\Delta[3.6\mu{m}] = 0$, as expected. The median of the distribution is $\Delta[3.6\mu{m}]=0.05$ with $\sigma_{\mathrm{MAD}}=0.20$.
For the reminder of this paper, we adopt $\sigma=0.20$ as the typical uncertainty on $\Delta[3.6\mu{m}]$.

\subsection{Dust attenuation}\label{subsec:sfr_rates}

To determine the intrinsic \lya \ flux we first need an estimate of the intrinsic \halpha \ flux.
The observed \halpha \ flux values are simply determined by multiplying \wha \ by the continuum flux derived from the best-fitting SED.
To dust-correct the observed fluxes we use the prescription of \citet{wuyts+13}:
\begin{equation}
    A_{\rm{H\,\alpha}, \rm{nebular}}=A_{\rm{H\,\alpha ,cont}}+0.9A_{\rm{H\,\alpha ,cont}}-0.15A_{\rm{H\,\alpha ,cont}}^2,
\end{equation}
where $A_{\rm{H\,\alpha ,cont}}$ is the continuum attenuation at $\lambda_{\rm{rest}}=6563$ {\AA} determined from the \textsc{fast++} SED fitting to the VANDELS DR4 photometry (see Section \ref{subsubsec:sed_fitting}).
The additional attenuation is physically motivated by the increased dust obscuration surrounding young star-forming regions \citep[e.g.,][]{wuyts+13, reddy2020}. We note that this conversion explicitly assumes a \citet{calzetti+00} attenuation law, which our previous work
has shown to provide a good description of the average dust attenuation in VANDELS galaxies down to stellar masses of $\rm{log(} M_{\star}/\rm{M_{\odot}})\simeq 9.0$ \citep{cullen+18}. We discuss the implications of assuming a steeper dust attenuation curve in Section \ref{sec:discussion}.

\begin{figure}
    \centerline{\includegraphics[width=1.05\columnwidth]{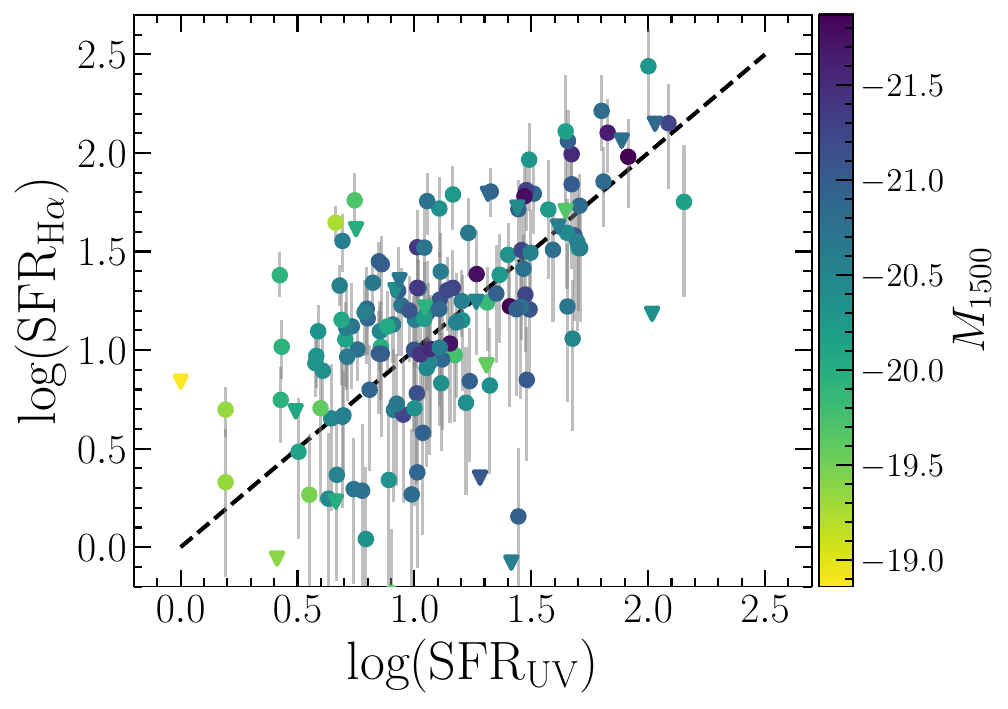}}
    \caption{A comparison of the star-formation rates derived from our dust-corrected {\halpha} and UV ($L_{1500}$) luminosities. The one-to-one relation is shown as the dashed line. The broad agreement between the two estimates demonstrates the relative robustness of the $L_{\rm{H\,\alpha}}$ measurements.}
    \label{fig:sfr_indicators}
\end{figure}

As a further sanity check on our approach we compare the star-formation rates inferred from the dust-corrected {\halpha} emission and dust-corrected FUV stellar continuum in Fig. \ref{fig:sfr_indicators}. Assuming that our SED-fitting and \wha \ estimates are robust, there should be good agreement between these two star-formation-rate indicators that are both sensitive to star-formation on $<100$ Myr timescales. To calculate the FUV-based star-formation rates we assume the \cite{madau+14} calibration
\begin{equation}
    {\rm{ SFR_{\rm{UV}}(M_{\odot}yr^{-1}) } = 6.58\times10^{-29}} L_{1500}(\rm{erg\ s^{-1} Hz^{-1}}),
    \label{eq:sfr_uv}
\end{equation}
where the dust-corrected $L_{1500}$ is calculated from the best-fitting SED template determined in Section \ref{subsubsec:sed_fitting} using a 100 \AA\ wide top-hat filter centered on $\lambda_{\rm{rest}}~=~1500$ {\AA}. To calculate the {{\halpha}}-based star-formation rate, we use the \citet[][]{reddy+18b} calibration
\begin{equation}
   {\rm{ SFR_{\rm{H\alpha}}(M_{\odot}yr^{-1}) } = 3.24\times10^{-42}}\ L_{\rm{H\alpha}}(\rm{erg\ s^{-1} }),
\end{equation}
where $L_{\rm{H\,\alpha}}$ is the dust-corrected \halpha\ luminosity. It can be seen from Fig. \ref{fig:sfr_indicators} that the 
two estimates are qualitatively in excellent agreement.
Fitting a fixed 1:1 relationship we find that the $\mathrm{log}(\rm{SFR_{H\,\alpha}})$ estimates are $+0.05$ dex ($\simeq 10\%$) larger than
the corresponding $\mathrm{log}(\rm{SFR_{UV}})$ estimates. However, we note that this systematic shift is much smaller than the statistical uncertainty on any individual measurement and globally decreasing the \halpha \ fluxes by $\simeq 10$ per cent does 
not affect any of the conclusions of this work.

\subsection{\lya \ escape fractions}\label{subsec:flya_constraints}

The \lya \ escape fraction is defined as the ratio between the observed and intrinsic \lya \ flux.
We determine intrinsic \lya \ fluxes from the dust-corrected \halpha \ fluxes under the assumption of Case-B recombination \citep[i.e., ${\mathrm{F}_{\rm{Ly\,\alpha ,int}}=8.7 \times \mathrm{F}_{\rm{H\,\alpha ,int}}}$;][see also; \citealt{henry+15}]{osterbrock_book} which yields,
\begin{equation}\label{eqn:flya}
f_{\rm{esc}}^{\rm{Ly\alpha}} = \frac{\mathrm{F}_{\rm{Ly\,\alpha ,obs}}}{8.7 \times \mathrm{F}_{\rm{H\,\alpha ,int}}}.
\end{equation}
For the objects in our sample with \wha\,$<0$ ($N=20$) we calculate a $3\sigma$ upper limit on \lyafesc.

\section{Results}\label{results}

\begin{figure*}
    \centerline{\includegraphics[width=2.0\columnwidth]{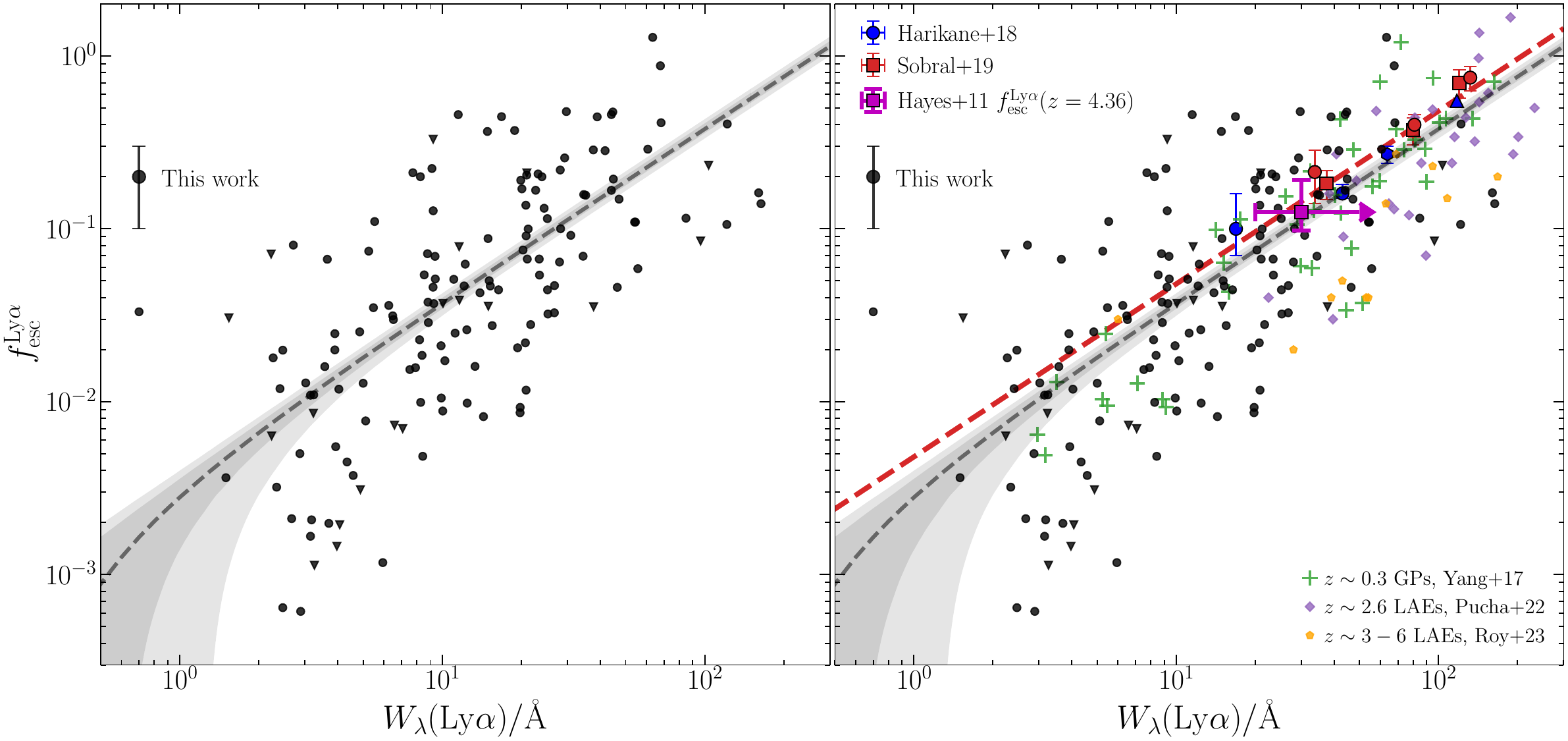}}
        \caption{The left-hand panel shows the relationship between $f_{\rm{esc}}^{\rm{Ly\alpha}}$ and {\wlya} for the final sample of $N=152$ VANDELS galaxies at $3.85\leq z_{\rm{spec}}\leq 4.85$ with {\wlya\,}$>0$ \AA. The best-fitting relation is shown as the dashed line and \lyafesc\, upper limits are shown as triangles (see text for details). The right-hand panel shows a comparison between our new results and
        those of previous literature studies. The best-fitting relation from \citet{sobral+19}, based on the combined sample of lower-redshift $z\sim0-0.3$ ‘high-redshift analogue’ galaxies and $z\sim2.2-2.6$ Lyman-alpha emitters (LAEs) (see \citealt{sobral+17} using stacks; and \citealt{trainor+15} using binning; red squares and circles, respectively), is shown as a red dashed line. A compilation of literature results based on LAE samples, including those from \citet{harikane+18} (LAE stacks, blue circles) at comparable redshifts ($z\sim4.9$), and constraints on individual $z\sim2.6$ (purple diamonds) and $z\sim3.0-6.0$ (orange pentagons) LAEs from \citet{pucha+22} and \citet{roy+23}, respectively, are also plotted. A sample of low-redshift ($z\sim0.3$) Green Pea galaxies \citep{yang+17} are shown as green crosses. The magenta square indicates the predicted $f_{\rm{esc}}^{\rm{Ly\alpha}}$ value at the median redshift of the sample, $\langle z_{\rm{spec}}\rangle\simeq4.36$, according to the $f_{\rm{esc}}^{\rm{Ly\alpha}}-z$ relation of \citet{hayes+11}.}
\label{fig:lyaew_lyafesc}
\end{figure*}
In this section we use our estimates of the observed \lya \ and \halpha \ line fluxes to place constraints on the \lya \ escape fraction and explore the relationship between \lyafesc \ and \wlya\,. At the end of the section we investigate whether the scatter around the \lyafesc$~-~$\wlya\ relation is in part driven by an underlying correlation between \lyafesc and \lycfesc.

\subsection{A non-evolving $\mathbf{f_{\rm{esc}}^{\rm{Ly\alpha}}-W_{\lambda}(\rm{Ly\,\alpha})}$ relation out to $\mathbf{z=5}$}
In the left-hand panel of Fig. \ref{fig:lyaew_lyafesc} we plot \lyafesc \ versus \wlya \ for our final sample of $N=152$ galaxies displaying \lya\ in emission, along with our best-fitting relation (dashed line).
Assuming the standard form of $f_{\rm{esc}}^{\rm{Ly\alpha}}=A \times W_{\lambda}(\rm{Ly\,\alpha}) + B$, we determine the best-fitting relation to be:
\begin{equation}
    f_{\rm{esc}}^{\rm{Ly\alpha}}=(3.8 \pm 0.3)\times10^{-3}W_{\lambda}(\rm{Ly\,\alpha})-(1.0 \pm 0.7)\times10^{-3},
\end{equation}
using the nested sampling algorithm \textsc{dynesty} \citep{speagle2020} with flat parameter priors.
Here, the best-fitting slope and intercept are given by the median of the posterior distribution, with the 1$\sigma$ errors designated as the $16^{\rm{th}}$ and $84^{\rm{th}}$ (68\%) percentiles. 
We find evidence for a correlation at the $\simeq 10\sigma$ level, with \lyafesc \ increasing monotonically from $\simeq 4\%$ to $\simeq 19\%$ over the range $10 \leq W_{\lambda}(\rm{Ly\,\alpha})\leq 50$ {\AA}. As expected, our best-fitting relation is consistent with \lyafesc$=0$ at \wlya$=0$.

In the right-hand panel of Fig. \ref{fig:lyaew_lyafesc} we compare our new results to those of previous studies in the literature. Our results are in good agreement with the relation (red dashed line) derived by \citet{sobral+19} for a combined sample of $z\simeq2.2-2.6$ \lya \ emitters and lower-redshift $z\sim0-0.3$ `high-redshift analogue' galaxies \citep[including Green Pea galaxies, LyC leakers, and {\halpha} emitters; e.g.,][]{hayes+13,henry+15,verhamme+17,yang+17}. Our results are also consistent with the stacking-based analysis of $N=99$ \lya \ emitters at a similar redshift to our sample ($z\simeq4.9$) presented by \citet{harikane+18}. It can be seen from Fig.~\ref{fig:lyaew_lyafesc} that our best-fitting relation has a somewhat lower normalisation than both the \citet{sobral+19} relation, and the binned \citet{harikane+18} data. However, the relations are consistent at the $< 2\sigma$ level and, given the different selection and analysis techniques applied in these other studies (i.e., stacking), we do not consider the offset to be significant. Indeed, the normalisation and scatter of our \emph{individual} measurements seem fully consistent with the $z=2.6$ \lya \ emitters presented in \citet{pucha+22} and the $z=0-0.3$ sample of Green Pea galaxies from \citet{yang+17}. We also show measurements from a sample of $z\simeq3.0-6.0\,$ LAEs analysed in \citet{roy+23}\, that are broadly consistent within the observed scatter. 

As discussed in \citet{sobral+19} \citep[see also][]{harikane+18} the observed {\lyafesc$-$\,\wlya} relation will be
influenced by $\xi_{\mathrm{ion}}$ and dust attenuation, both of which are linked to metallicity. Given that these 
properties are known to evolve with redshift \citep{emani+16,matthee+16}, the apparent non-evolution of the {\lyafesc$-$\,\wlya} relation shown in Fig. \ref{fig:lyaew_lyafesc} is worthy of further consideration.

Crucially, the $z\sim0-0.3$ galaxy samples we compare to in Fig.~\ref{fig:lyaew_lyafesc} have been deliberately selected to be analogues of high-redshift SFGs, such as those comprising our $z\simeq 4-5$ sample from VANDELS.
Using follow-up rest-frame optical spectroscopy of $N=33$ VANDELS SFGs at $z\gtrsim3$, \citet{cullen+21} measured metallicites spanning $12+\mathrm{log(O/H)}\simeq7.6-8.2$, comparable to the range displayed by the \citet{yang+17} Green Pea galaxy sample \citep[also included in the $z\sim0.3$ sub-sample of][]{sobral+19}. We can further deduce that the low-redshift samples and our own $z\simeq4-5$ sample will have comparable $\xi_{\mathrm{ion}}$ as a result of their similar metallicities \citep{cullen+20}. Consequently, the lack of evolution in the {\lyafesc$-$\,\wlya} relation shown in Fig. \ref{fig:lyaew_lyafesc}, simply indicates that the physical processes regulating the production and 
escape of \lya\ photons in high-redshift SFGs are comparable to those in low-redshift analogues, deliberately selected to
be a close match in terms of metallicity, $\xi_{\mathrm{ion}}$ and dust attenuation.

We note that, due to the depth of the VANDELS spectroscopy, we are able to trace the \lyafesc$-$\,\wlya \ relation for \emph{individual} objects for the first time at these redshifts, as well as extending the relation to \wlya$\lesssim 20$ {\AA}, a regime previously only accessible to low-redshift studies.
Our analysis demonstrates that at $z\geq 4$ the \lyafesc$-$\,\wlya \ relation extends down to \wlya\,$\simeq 2-3$ \AA, with our \lyafesc\ constraints at these low \wlya \ values directly comparable to those derived in the low-redshift Universe (Fig.~\ref{fig:lyaew_lyafesc}).

Overall, our analysis indicates that the ${f_{\rm{esc}}^{\rm{Ly\alpha}}}-W_{\lambda}(\rm{Ly\,\alpha})$ relation for SFGs at $z\simeq4-5$ is indistinguishable from that followed by their low-redshift analogues. This implies that the physical processes determining the production and escape of {\lya} photons from low metallicity, high $\xi_{\mathrm{ion}}$ galaxies do not vary significantly over $\simeq 11$ Gyr (i.e., $\simeq 90$ per cent) of cosmic time. 

\begin{figure*}
    \centerline{\includegraphics[width=2\columnwidth]{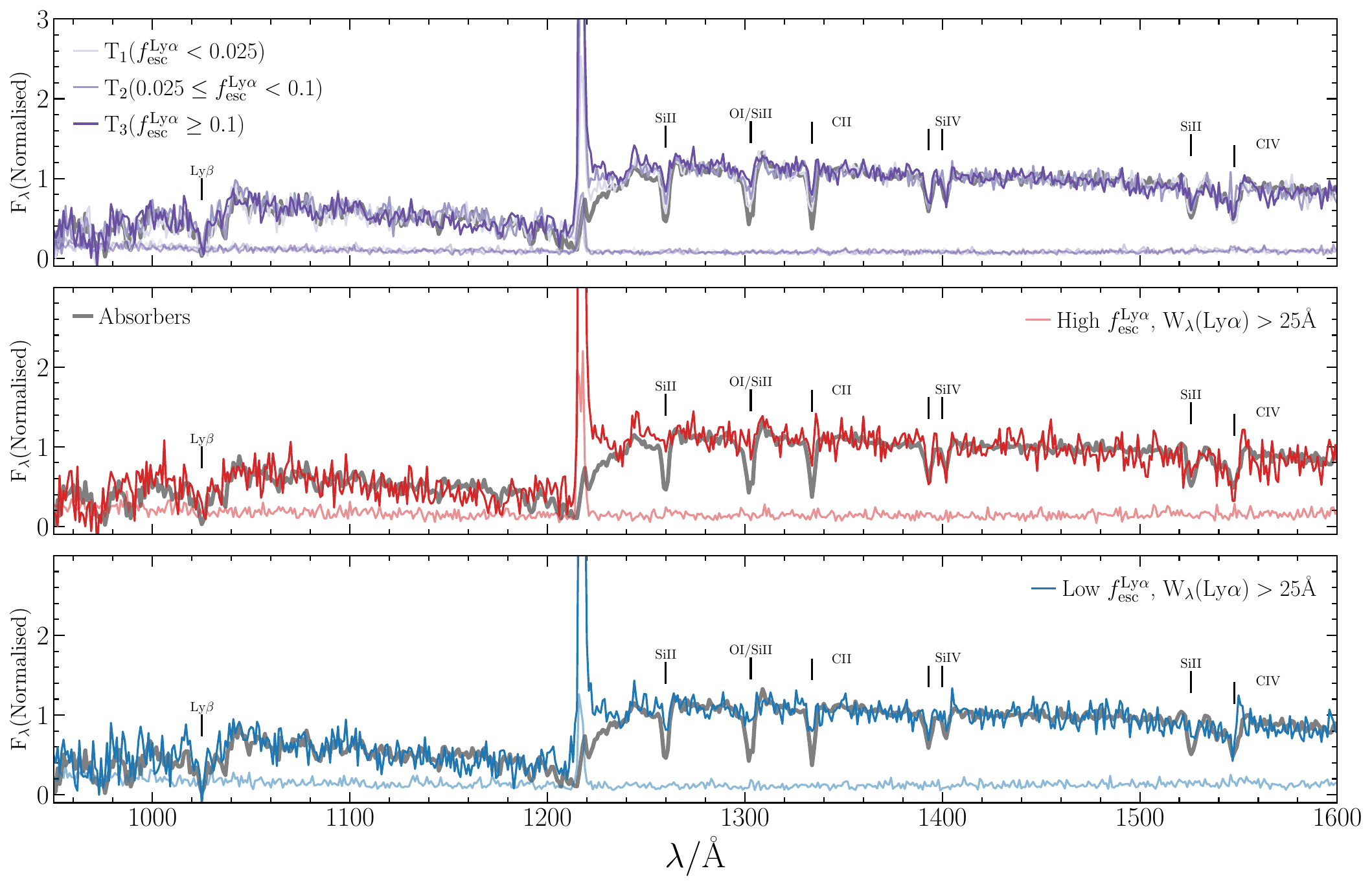}}
    \caption{The top panel shows an overlay of four composite VANDELS spectra. The dark grey composite is a stack of $N=111$ VANDELS 
    galaxies within the redshift range $3.85\leq z_{\rm spec}\leq 4.95$ that display {\lya} in absorption (i.e. $W_{\rm{Ly\,\alpha}}<0$). This composite is shown in all three panels. The purple (light to dark) composites are constructed from three equally-occupied bins of \lyafesc\, (T$_{1}:$\lyafesc$<0.025$, T$_{2}:0.025\leq$\,\lyafesc$<0.1$ and T$_{3}:$\lyafesc$\geq 0.1$). The middle and bottom panels show composites formed from objects with $W_{\rm{Ly\,\alpha}}\geq$ 25 \AA, that have been split into high- and low-\lyafesc sub-samples at a threshold of \lyafesc$\geq0.2$. This selection ensures that both composites have approximately the same equivalent width ($\Delta(W_{\lambda}(\rm{Ly\,\alpha}))<5$ {\AA}) but widely different values of \lyafesc\,(see text for details). The error spectra are shown in light colours and notable absorption features are highlighted with black vertical lines (e.g. Si{\sc{ii}$\lambda$1260}, C{\sc{ii}$\lambda$1334}, O{\sc{i}}/Si{\sc{ii}}{$\lambda$1303}, and Si{\sc{ii}$\lambda$1526}).}
    \label{fig:spec_plot_wide}
\end{figure*}

\subsubsection{Average \lyafesc \ of \lya \ emitters}

It is interesting to compare our results for individual objects to previous constraints on the population averaged \lya \ escape fraction.
Based on a compilation of \lya, UV, and \halpha \ luminosity functions, \citet{hayes+11} determined that the redshift evolution of $\langle f_{\rm{esc}}^{\rm{Ly\alpha}}\rangle$ is well-described by a power law of the form $\propto (1+z)^{2.57}$, with $\langle f_{\rm{esc}}^{\rm{Ly\alpha}}\rangle$ evolving from $\langle f_{\rm{esc}}^{\rm{Ly\alpha}}\rangle\simeq0.01$ at $z\simeq0.3$ up to $\langle f_{\rm{esc}}^{\rm{Ly\alpha}}\rangle\simeq0.4$ at $z\simeq6$.
At the median redshift of our final sample ($\langle z_{\rm{spec}}\rangle = 4.36$), the \citet{hayes+11} relation predicts $\langle f_{\rm{esc}}^{\rm{Ly\alpha}}\rangle \simeq 0.13$ which is indicated by the purple data point in the right-hand panel of Fig. \ref{fig:lyaew_lyafesc}.

This average value is clearly in good agreement with the high \wlya \ objects in our sample, consistent with the fact that the galaxies used to derive the \citet{hayes+11} relation were typically \lya \ emitters with \wlya\,$>20$ {\AA}.
Restricting our sample to objects with \wlya\,$>20$ {\AA} we find a median value of $\langle f_{\rm{esc}}^{\rm{Ly\alpha}}\rangle\simeq0.12$, in excellent agreement with the \citet{hayes+11} prediction. This consistency is encouraging, given that the two estimates originate from completely independent methods.

\subsection{Connecting \lya \ and LyC escape}\label{sec:lycfesc}
Although the results presented in Fig. \ref{fig:lyaew_lyafesc} show a strong correlation between \lyafesc\ and \wlya, there is clearly a large amount of associated scatter ($\simeq 0.5$ dex). Some fraction of this scatter is attributable to measurement uncertainties, given that our individual estimates of \lyafesc\ are undoubtedly noisy. However, at a given value of \wlya\, we also expect {\it intrinsic} scatter in \lyafesc\ due to variations in the stellar populations and dust/gas properties. One way of exploring the range of \lyafesc at a given value of \wlya\, is to investigate the link between \lyafesc and \lycfesc.

A growing body of literature has empirically established a strong positive correlation between \wlya \ (i.e., \lyafesc) and\lycfesc in high-redshift star-forming galaxies, either via direct measurements of LyC emission \citep[e.g.,][]{marchi+18,steidel+18,pahl+21,begley+22}, or via indirect studies characterising the H{\mbox{\,\sc{i}} covering fraction \citep{shapley+03,gazagnes+20,saldana-lopez+22b}. 
A similarly strong connection is also observed at low redshifts \citep[e.g.,][]{vanzella+16,verhamme+17,flury+22b}. 
This link can be explained by the similar escape path of LyC and \lya \ photons through low dust/H{\mbox{\,\sc{i}}} column-density channels \citep[e.g.,][]{dijkstra+16,jaskot+19}, a picture supported by detailed radiative transfer simulations \citep[e.g.,][]{cen+15,kimm+19}.

Below, we use composites of our VANDELS spectra to explore this connection further.
Rather than using \wlya \ as a proxy for \lyafesc, we can take advantage of our \emph{individual} \lyafesc \ escape estimates to trace the connection between \lyafesc \ and the properties of the ISM measured from deep rest-frame FUV spectra.
This in turn allows us to connect the escape of \lya \ and LyC photons using the correlation established between the equivalent width of low-ionization-state FUV ISM absorption lines and \lycfesc\, by \citet{saldana-lopez+22b}.

\subsubsection{Constructing VANDELS composite spectra}

To maximise the available signal-to-noise of our VANDELS spectra we create stacked FUV composites following a similar procedure to that outlined in \citet{cullen+19}. 
To create the composite spectra of a given ensemble of galaxies, we first shift each individual spectrum to the rest-frame using its spectroscopic redshift and normalise to the median flux in the range $1420\leq\lambda_{\rm{rest}}\leq1480$ \AA. 
The individual flux elements of each spectra are then binned onto the desired wavelength grid of the final stack (1 {\AA}/pix). The flux of each pixel in the composite spectrum is given by the median of the individual fluxes after $3\sigma$ outliers have been sigma clipped. 
The associated error spectrum is estimated by bootstrap re-sampling of the fluxes in each wavelength bin, taking the standard deviation of each as the $1\sigma$ uncertainty.

\begin{figure*}
    \centerline{\includegraphics[width=2.0\columnwidth]{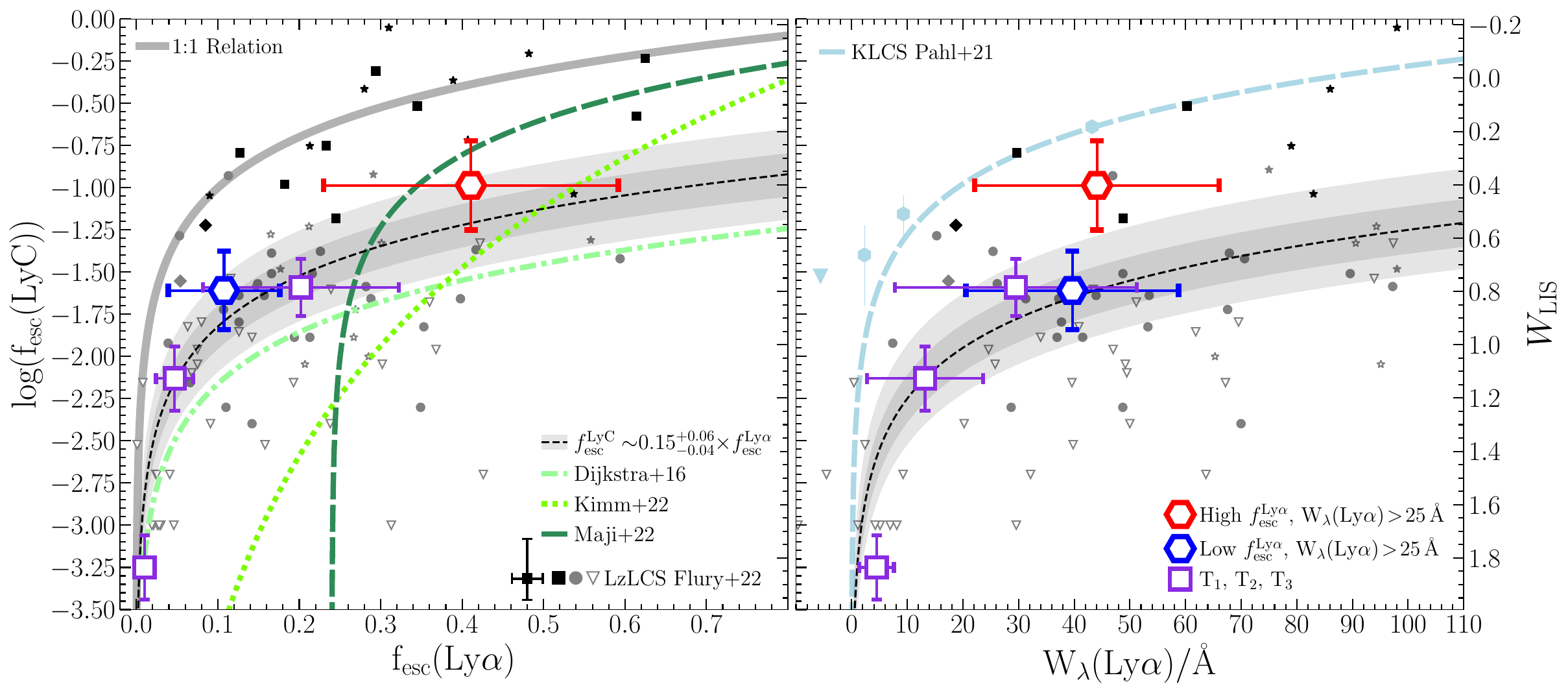}}
    \caption{The left-hand panel shows the relationship between \lycfesc\, and \lyafesc\, based on the composite spectra shown in Fig.~\ref{fig:spec_plot_wide}, where \lycfesc\, has been inferred from $W_{\rm{LIS}}$ measurements (shown on the secondary y-axis on the right-hand side) using the $W_{\rm{LIS}}-$\lycfesc relation from \citet{saldana-lopez+22b}. The three composites based on equally-occupied bins of \lyafesc\, are shown as purple squares, while the high- and low-\lyafesc composites at a constant value of \wlya\, are shown as red and blue hexagons. The \lyafesc\ and \lycfesc\ data from the LzLCS sample for strong Lyman continuum emitters (LCEs), weak LCEs and non-emitters are shown as small filled black, filled grey and open grey markers, respectively \citep[typical uncertainties are shown in the lower right, see][for further details]{flury+22a,flury+22b}. Simulation-derived \lyafesc$-$\lycfesc relations are shown as green lines (\citealt{dijkstra+16}, dash-dotted; \citealt{kimm+22}, dotted; \citealt{maji+22}, dashed). The 1:1 relation is shown as thick grey line and \lycfesc$\simeq 0.15^{+0.06}_{-0.04}\times$\lyafesc is shown as a dashed black line. The right-hand panel shows the same composite-based \lycfesc\, constraints as a function of \wlya, including values for the LzLCS sample. The {\wlya$-$\lycfesc} relation derived at $z\simeq 3$ from the KLCS \citep{pahl+21} is shown as the dashed light-blue line, with the light-blue markers indicating constraints for composite KLCS spectra as a function of {\wlya}. The dashed black line shows the relationship \lycfesc$\simeq 0.0005$\lyafesc, derived by combining our best-fitting \lyafesc$-$\wlya\, relationship (see Fig.~\ref{fig:lyaew_lyafesc}) with \lycfesc$\simeq0.15^{+0.06}_{-0.04}\times$\lyafesc.}
    \label{fig:fescLyc_vs_lya_combined}
\end{figure*}

\subsubsection{The strength of ISM absorption features}\label{subsec:composite_stacks}

The top panel of Fig. \ref{fig:spec_plot_wide} is an overlay of four composite spectra. The dark-grey spectrum is a composite of the $N=111$ VANDELS galaxies within the redshift range $3.85\leq z_{\rm spec} \leq 4.95$ that display \lya\ in absorption. The other three spectra are composites obtained by splitting our final sample of $N=152$ VANDELS galaxies with \lya\ in emission into three equally occupied bins of \lyafesc\, (T$_{1}=$\lyafesc$<0.025$, T$_{2}=0.025\leq$\lyafesc$<0.1$ and T$_{3}=$\lyafesc$\geq 0.1$).

Some key features are immediately visible by eye. Compared to the composite spectrum of galaxies displaying \lya\, in absorption, it is clear that the {\lya} emission composites display progressively weaker low-ionization-state ISM absorption lines as a function of increasing \lyafesc. The weak LIS ISM lines are a clear signature of a low covering fraction\footnote{We highlight that the LIS lines are saturated and therefore act as tracers of the covering fraction, $C_{f}$ (see also Section \ref{subsubsec:lyafesc_igm_state}). For the Si{\mbox{\,\sc{ii}}} line species, we measure $W_{1260}({\rm{Si}}{\mbox{\,\sc{ii}}})/W_{1526}({\rm{Si}}{\mbox{\,\sc{ii}}})\lesssim 1.6\pm0.5$ for all three {\lyafesc}-binned composites (top panel, Fig. \ref{fig:spec_plot_wide}). In the optically-thin regime $W_{1260}({\rm{Si}}{\mbox{\,\sc{ii}}})/W_{1526}({\rm{Si}}{\mbox{\,\sc{ii}}})=6.0$ \citep{shapley+03,erb+10}, which is inconsistent with the data.} of neutral \hi \ gas, one of the key requirements for an increased escape of \lya/LyC photons \citep{reddy+16, saldana-lopez+22a}.

Although the LIS ISM features in the composite spectra shown in Fig. \ref{fig:spec_plot_wide} behave as expected, the correlation between \lyafesc\, and \wlya\, shown in Fig.~\ref{fig:lyaew_lyafesc} means that it is not clear which of the two parameters is driving the observed trend. To investigate this issue, we also create two composite spectra at \emph{fixed} \wlya. By selecting all galaxies with $W_{\lambda}(\rm{Ly\,\alpha})\geq25$ {\AA} and splitting into high-\lyafesc\ and low-\lyafesc\ sub-sets at a threshold of \lyafesc$\geq0.2$, it was possible to produce two composites with similar \wlya\, and sufficiently high signal-to-noise to allow measurements of the LIS absorption features.
 The high-\lyafesc\ and low-\lyafesc\ composites are shown in the middle and bottom panel of Fig. \ref{fig:spec_plot_wide}, respectively. The high-\lyafesc composite contains $N=13$ galaxies with median (and $\sigma_{\mathrm{MAD}}$) values of $\langle f_{\rm{esc}}^{\rm{Ly\alpha}}\rangle=0.41\pm0.18$ and  $\langle W_{\lambda}(\rm{Ly\,\alpha})\rangle=44.1\pm21.9$ {\AA}. 
The low-\lyafesc composite contains $N=22$ galaxies with $\langle f_{\rm{esc}}^{\rm{Ly\alpha}}\rangle=0.11\pm0.07$ and $\langle W_{\lambda}(\rm{Ly\,\alpha})\rangle=39.7\pm19.1$ {\AA}. 

\subsubsection{LIS equivalent widths}
For each spectrum, we measure equivalent widths for the Si{\mbox{\,\sc{ii}}{$\lambda1260$}}, C{\mbox{\,\sc{ii}}{$\lambda1334$}}, and the O{\mbox{\,\sc{i}}{$\lambda1303$}}$+$Si{\mbox{\,\sc{ii}}{$\lambda1303$}} low-ionization state ISM absorption features (with the final feature blended due to the VANDELS $R\simeq600$ spectral resolution; see Fig. \ref{fig:spec_plot_wide}).
The equivalent widths are calculated numerically according to the following equation:
\begin{equation}
    W_{\rm{LIS}}=\int_{\Delta\lambda}\left( 1 - \frac{f_{\rm{obs}}}{f_{\rm{cont}}} \right)d\lambda,
\end{equation}
where $f_{\rm{obs}}$ is the flux density of observed spectrum, $f_{\rm{cont}}$ is the underlying stellar continuum flux density, and $\Delta\lambda$ is the width of the region over which the numerical integration is performed, which we set to a default value of $\pm500\,\mathrm{km\,s}^{-1}$.
The values of $\Delta\lambda$ and $\lambda_{0}$ are manually adjusted for each line measurement to account for velocity offsets, nearby noise spikes or potential non-resonant fine structure emission\footnote{Due to the relatively low resolution of the VANDELS spectra, it is not possible to fully assess the impact of absorption line infilling from non-resonant fine structure emission in the vicinity of the LIS features. However, the close agreement of the observed {{\wlya}$-W_\mathrm{LIS}$} relation with other literature measurements of SFGs with higher resolution spectra \citep[e.g.,][]{shapley+03,du+18,pahl+20}, provides confidence that our $W_\mathrm{LIS}$ measurements are not significantly influenced by infilling.}. 
The stellar continuum component is calculated by fitting a $f_{\lambda}\propto\lambda^{\beta}$ power law to the continuum either side of the line, typically spanning $\pm4500\,\mathrm{km\,s}^{-1}$ ($\gtrsim20$ \AA).
Uncertainties were calculated using a Monte Carlo procedure.

\subsubsection{The connection between \lyafesc and the ISM ionization state}\label{subsubsec:lyafesc_igm_state}
Qualitatively, it can be seen from Fig. \ref{fig:spec_plot_wide} that the low-ionization ISM lines are somewhat weaker in the high-\lyafesc \ composite than in the low-\lyafesc \ composite. In addition, the high-ionization Si{\mbox{\,\sc{iv}}}$\lambda\lambda1393,1402$ and blended C{\mbox{\,\sc{iv}}}$\lambda\lambda1548,1550$ absorption lines appear to be more visually prominent in the high-\lyafesc \ composite. That is, at fixed \wlya, galaxies with higher \lyafesc\ appear to show signatures consistent with having a lower covering fraction of neutral gas and a higher covering fraction of ionized gas. 

Using the procedure outlined above, we measure equivalent widths for the Si{\mbox{\,\sc{iv}}}$\lambda\lambda1393,1402$ doublet in the high-{\lyafesc} composite of $W_{1393}({\rm{Si}}{\mbox{\,\sc{iv}}})=1.36\pm0.33$ {\AA} and $W_{1402}({\rm{Si}}{\mbox{\,\sc{iv}}})=1.23\pm0.34$ {\AA}. Similarly, in the low-{\lyafesc} composite we measure $W_{1393}({\rm{Si}}{\mbox{\,\sc{iv}}})=0.74\pm0.31$ {\AA} and $W_{1402}({\rm{Si}}{\mbox{\,\sc{iv}}})=0.51\pm0.20$ {\AA}. 

The line ratio in the Si{\mbox{\,\sc{iv}}}$\lambda\lambda1393,1402$ doublet can be used to infer whether it originates from optically thin or optically-thick ISM gas. In the optically-thin regime, the doublet ratio will be $W_{1393}({\rm{Si}}{\mbox{\,\sc{iv}}})/W_{1402}({\rm{Si}}{\mbox{\,\sc{iv}}}){\simeq2}$ \citep{shapley+03,berry+12}, whereas ratios of $W_{1393}({\rm{Si}}{\mbox{\,\sc{iv}}})/W_{1402}({\rm{Si}}{\mbox{\,\sc{iv}}})\simeq1$ are indicative of saturated lines arising from an optically thick, high-ionization-state ISM. 
For the high- and low-{\lyafesc} composites we find doublet ratios of $W_{1393}({\rm{Si}}{\mbox{\,\sc{iv}}})/W_{1402}({\rm{Si}}{\mbox{\,\sc{iv}}})=1.1\pm0.4$ and $W_{1393}({\rm{Si}}{\mbox{\,\sc{iv}}})/W_{1402}({\rm{Si}}{\mbox{\,\sc{iv}}})=1.45\pm0.8$, respectively. 
For the low-{\lyafesc} composite we are clearly unable to conclude anything regarding optical depth, as the doublet ratio is fully consistent with both the optically thin and optically-thick regimes. On the other hand, despite the obvious uncertainty, the high-{\lyafesc} doublet ratio is more consistent with saturation, providing evidence of a more-highly ionized ISM environment in the galaxy composite with higher {\lyafesc}.

This conclusion is further strengthened by comparing the UV spectral slopes of the high- and low-{\lyafesc} composites. Fitting a power-law of the form $f_{\lambda}\propto\lambda^{\beta}$, we find the high-{\lyafesc} composite ($\beta=-1.8\pm0.2$) to be bluer than its low-{\lyafesc} counterpart ($\beta=-1.4\pm0.1$). This spectral slope difference is consistent with recent results \citep[see][]{gazagnes+20,begley+22} indicating that galaxies with bluer UV slopes are more likely to display more-highly ionizing environments \citep[and higher \lycfesc;][]{chisholm+22}.

\subsubsection{The connection between \lyafesc\, and \lycfesc}\label{subsubsec:flyc_method}
Recently, \citet{saldana-lopez+22b} have identified a strong relationship between \lycfesc and the equivalent width of low-ionization absorption lines ($W_{\rm{LIS}}$), calibrated using the LzLCS dataset \citep[see also][]{chisholm+18,gazagnes+20}. 
Using this relationship, we are able to estimate the value of \lycfesc \ for the composite spectra shown in Fig.~\ref{fig:spec_plot_wide}.

We calculate the inverse-variance weighted mean across the Si{\mbox{\,\sc{ii}}{$\lambda1260$}}, C{\mbox{\,\sc{ii}}{$\lambda1334$}}, O{\mbox{\,\sc{i}}{$\lambda1303$}}$+$Si{\mbox{\,\sc{ii}}}{$\lambda1303$}} and the Si{\mbox{\,\sc{ii}}{$\lambda1526$}} features\footnote{These lines represent a subset of the full suite of LIS features originally employed in \citet{saldana-lopez+22b}. Specifically, we do not include measurements of the Si{\mbox{\,\sc{ii}}{$\lambda989$}}, Si{\mbox{\,\sc{ii}}{$\lambda1020$}}, and Si{\mbox{\,\sc{ii}}{$\lambda\lambda1190,1193$}} features and in turn avoid potential systematic biases that may arise from the need to correct for the impact of IGM+CGM absorption blueward of {\lya}. We also exclude the  Si{\mbox{\,\sc{ii}}{$\lambda1526$}} absorption feature from the high-\lyafesc\ composite spectra due to a suspected noise spike in the immediate wavelength vicinity affecting the measurement.}. For the three composite spectra shown in the upper panel of Fig.~\ref{fig:spec_plot_wide}, we measure values of $W_{\rm{LIS}}=1.84\pm0.12$ {\AA}, $1.13\pm0.12$ {\AA} and $0.78\pm0.17$~{\AA}, in order of increasing \lyafesc. Likewise, for the high-\lyafesc \ and low-\lyafesc \ composites shown in the middle and bottom panels of Fig.~\ref{fig:spec_plot_wide}, we measure $W_{\rm{LIS}}=0.40\pm0.17$ {\AA} and $0.80\pm0.15$\,\AA, respectively. We show the corresponding values of \lycfesc in the left-hand panel of Fig.~\ref{fig:fescLyc_vs_lya_combined}, having employed Eqn. 11 from \citet{saldana-lopez+22b} to map between $W_{\rm{LIS}}$ and \lycfesc. For galaxies at $z\simeq4-5$, our results indicate that \lyafesc \ and \lycfesc\, are positively correlated and follow a relationship of the form  \lycfesc$\simeq 0.15^{+0.06}_{-0.04}$\lyafesc. 
Crucially, we can be confident that the \lycfesc$-$\lyafesc relation is not being driven by \wlya, given that the \lycfesc\, and \lyafesc\, values derived for the two equal-\wlya\, composites follow the same relation (red and blue hexagons in Fig.~\ref{fig:fescLyc_vs_lya_combined}). We discuss the physical interpretation and implications of this result in Section 5.

\subsubsection{Comparing \lycfesc \ inferred from $W_{\rm{LIS}}$ to direct \lycfesc \ observations at $z>3$}
Finally, in the right-hand panel of Fig.~\ref{fig:fescLyc_vs_lya_combined} we compare the \lycfesc$-$\wlya\, relationship of our $z\simeq4-5$ sample, as estimated from LIS absorption-line strength, to more direct measurements at low and high redshift. Our results are consistent with a relation of the form \lycfesc$\simeq 0.0005$\wlya\, (black dashed line), which is simply the result of combining the \lyafesc$-$\wlya\, relation shown in Fig.~\ref{fig:lyaew_lyafesc} with the \lycfesc$-$\lyafesc\, relation shown in the left-hand panel of Fig.~\ref{fig:fescLyc_vs_lya_combined}.
As expected, given the origin of the $W_{\rm{LIS}}-$\lycfesc calibration from \citet{saldana-lopez+22b}, our results are consistent with the \lycfesc$-$\,\wlya\, measurements from the low-redshift LzLCS survey at $z=0.2-0.4$ \citep{flury+22a}, both in terms of normalisation and scatter.

In contrast, our new results are systematically lower than the \lycfesc$-$\,\wlya\, relation derived by \citet{pahl+21}. The \wlya$-$\lycfesc \ constraints from \citet{pahl+21} are based on direct measurements of \wlya \ and LyC flux from extremely deep spectra of galaxies at $z\simeq3$ (KLCS survey; \citealt{steidel+18})
and correspond to \lycfesc$\simeq 0.0053$\wlya\, (blue dashed line). Although our new results at $z\simeq4-5$ follow a \lycfesc$-$\wlya\, relation with the same functional form, the use of the low-redshift $W_{\rm{LIS}}-$\lycfesc calibration from \citet{saldana-lopez+22b} leads to a normalisation that is a factor of $\simeq 10$ lower. 

To confirm the presence of this normalisation discrepancy, we construct an additional composite VANDELS spectrum from the upper-\wlya \ sub-sample defined in \citet{begley+22} (with SFGs in the redshift range $3.35<z_{\mathrm{spec}}<3.95$ and \wlya\, spanning $-6${\,\AA\,}$\lesssim W_{\lambda}(\mathrm{Ly\,\alpha})\lesssim 93${\,\AA}), and infer {\lycfesc}{$\simeq0.005$} (corresponding to $W_{\mathrm{LIS}}\simeq1.2$\,{\AA}), following the same method detailed in Section \ref{subsubsec:flyc_method}. Comparing with the direct photometry-based constraint of \lycfesc{$=0.12^{+0.06}_{-0.04}$} measured in \citet{begley+22}, we again find independent evidence for a significant (factor $\gtrsim10-20$) normalisation offset at $z\geq3$ between indirect {\lycfesc} estimates (using the low-redshift {\lycfesc}$-\,W_{\mathrm{LIS}}$ calibration) and direct estimates of {\lycfesc} from deep \textit{U-}band imaging/spectroscopy.

To achieve consistency, it seems likely that the normalisation of the $W_{\rm{LIS}}-$\lycfesc relation must evolve with redshift to permit significantly higher values of \lyafesc\, for a given value of $W_{\rm{LIS}}$. The offset in normalisation was previously noted and discussed in detail by \citet{saldana-lopez+22b}. In Section \ref{sec:discussion} we provide a brief review of the systematics that are likely to be responsible.

\section{Discussion} \label{sec:discussion}

The results presented in Fig. \ref{fig:lyaew_lyafesc} demonstrate that our sample of $z\simeq4-5$ galaxies displays a \lyafesc$-$\wlya\, relation that is very similar to the relation observed at lower redshifts. We also show in Fig. \ref{fig:spec_plot_wide} that some of the scatter around this relationship is connected to a genuine range in {\lyafesc} at a given value of {\wlya}. This figure demonstrates that empirically,
composite spectra binned as a function of \lyafesc\, display the expected anti-correlation between \lyafesc\, and the depth of the LIS absorption features that are believed to 
be tracers of \lycfesc (and \lyafesc). Based on a low-redshift calibration between $W_{\rm{LIS}}$ and \lycfesc from \citet{saldana-lopez+22b}, the results presented in the left-hand panel of Fig.~\ref{fig:fescLyc_vs_lya_combined} show that \lycfesc\, and \lyafesc\, are strongly correlated, following a relation consistent with \lycfesc$\simeq0.15^{+0.06}_{-0.04}$\lyafesc.
Importantly, the results derived from the composite spectra designed to have the same \wlya\, indicate that this correlation is not being driven by a secondary correlation between \lyafesc\, and \wlya.

These results are qualitatively consistent with the predictions of so-called `picket-fence' or `holes' models, in which the geometry and physical conditions of the ISM gas and dust in the immediate vicinity of star-forming regions play a decisive role in governing the escape of ionizing and {\lya} photons \citep{shapley+03,chisholm+18,gazagnes+20}. These models suggest \lycfesc is primarily dictated by the covering fraction of optically thick H{\mbox{\,\sc{i}}} gas  ($C_{f}^{\rm{H{\mbox{\,\sc{i}}}}}$), with LyC and \lya\, photons escaping through channels of low-column-density and/or high-ionization-state gas. These channels become more abundant with decreasing $C_{f}^{\rm{H{\mbox{\,\sc{i}}}}}$, leading to a \lycfesc$\propto (1-C_{f}^{\rm{H{\mbox{\,\sc{i}}}}})$ relationship \citep{verhamme+17,steidel+18,saldana-lopez+22b}.

Our results are, to first order, also consistent with the overall trends expected from simulations.
In the left-hand panel of Fig. \ref{fig:fescLyc_vs_lya_combined} we show the predictions of three independent simulations.
The dashed green line shows the \lyafesc$-$\lycfesc \ relation from \citet{maji+22}, based on the SPHINX suite of cosmological radiation-hydrodynamical simulations \citep{rosdahl+18}.
The dash-dotted green line shows the relation from \citet{dijkstra+16}, derived from a suite of clumpy ISM models covering a wide range of physical conditions.
Finally, the dotted green line shows the relation from the \citet{kimm+22} high-resolution simulations of giant molecular clouds, including stellar feedback, that suggest a steeper relationship of the form {\lycfesc$\simeq (f_{\rm{esc}}^{\rm{Ly\alpha}})^{3.7}$}. 
All three studies show trends qualitatively consistent with our results, albeit with offsets in absolute {\lycfesc} values. The \citet{maji+22} relation appears to imply that a threshold of {\lyafesc$\simeq0.2$} must be met before LyC leakage can occur, inconsistent with our low-\lyafesc \ constraints. However, \citet{maji+22} also show that a number of their simulated galaxies do occupy this region of \lyafesc$-$\lycfesc parameter space and that there is generally a large amount of scatter in the relation, consistent with the low-redshift observations.

A final issue that merits discussion is the offset in absolute \lycfesc\,inferred indirectly from low-ionization FUV absorption lines and direct estimates of \lycfesc at $z>3$ \citep{steidel+18,pahl+21, begley+22}. 
This issue has been explored in detail by \citet{saldana-lopez+22b} and we refer the reader to that paper for a thorough discussion. However, there are three obvious systematic effects that are worth briefly discussing here. 

Firstly, in general, it is worth remembering that the $W_{\rm{LIS}}-f_{\rm{esc}}^{\rm{LyC}}$ relation used here was calibrated using a sample of rare, low-redshift LCE candidates from the LzLCS \citep{saldana-lopez+22b}. These galaxies will not be perfect analogues for the $z>3$ star-forming population in terms of evolutionary stage, star-formation rate, metallicity or dust enrichment. As such, it is not unreasonable to think that the appropriate absolute calibration of the $W_{\rm{LIS}}-f_{\rm{esc}}^{\rm{LyC}}$ at $z>3$ may well be significantly different from that which is appropriate at low redshifts.

A second potentially important systematic is the choice of model used to describe the ISM gas geometry. 
In the `holes' geometry adopted by \citet{pahl+21}, the dust and H{\mbox{\,\sc{i}}} gas reside in pockets around star-forming regions and are optically thick to ionizing radiation. 
In this geometry, the LyC escape fraction is simply given as {\lycfesc$=(1-C_{f}^{\rm{H{\mbox{\,\sc{i}}}}})$}, where $C_{f}^{\rm{H{\mbox{\,\sc{i}}}}}$ is the covering fraction of neutral hydrogen gas.
Recently, \citet{chisholm+18} have suggested that this model, by not considering the effects of dust outside the optically thick gas clumps, typically overestimates {\lycfesc} \citep[e.g., see also][]{gazagnes+20,kakiichi+gronke+21}. 
Accounting for this could potentially lower the {\lycfesc} estimates presented in \citet{pahl+21}.
However, we note that the {\lycfesc$=(1-C_{f}^{\rm{H{\mbox{\,\sc{i}}}}})$} approximation should be valid at the high \wlya\, values we are considering here, since these galaxies are generally expected to have low dust attenuation. Indeed, \citet{saldana-lopez+22b} conclude that this geometric consideration is insufficient to reconcile the \lycfesc \ estimates.

Finally, it is clear that the choice of dust attenuation law can have a strong systematic impact on the derived values of \lycfesc \citep[e.g.,][]{begley+22}. Indeed, \citet{saldana-lopez+22b} note that their $W_{\rm{LIS}}-f_{\rm{esc}}^{\rm{LyC}}$ calibration would predict \lycfesc\, values that are higher by up to a factor 1.5, if they switch from the \cite{reddy+16} dust attenuation law to a steeper SMC-like law \citep[e.g.,][]{gordan+03}, which would be closer to the assumptions 
employed by \cite{pahl+21} at high \wlya. However, even in this case, the offset between the two estimates of \lycfesc\, would still be a factor of $\simeq 6$.

Ultimately, a combination of all of these different factors is likely to be having an impact. However, it is worth noting that, although the {\it absolute} \lycfesc \ values are subject to potentially large systematic uncertainties, the {\it relative} values should be much less affected. Consequently, our conclusion that \lyafesc\, and \lycfesc are strongly correlated, and that the correlation is not driven by varying \wlya, should still be robust, regardless of the 
exact normalization of either value.

\section{Summary}\label{sec:summary}
We have presented the results of a study exploring the connection between the Ly$\alpha$ escape fraction ({\lyafesc}) and the Lyman continuum escape fraction ({\lycfesc}) for a sample of $N=152$ SFGs selected from the ESO VANDELS spectroscopic survey \citep{mclure+18,pentericci+18,garilli+21} at {$3.85\leq z_{\rm{spec}}\leq4.95$}. 

We combine measurements of {\wlya} from ultra-deep, rest-frame FUV VANDELS spectra with {\halpha} equivalent widths derived from IRAC $3.6\,\mu\mathrm{m}$ flux-excess measurements to estimate \emph{individual} {\lyafesc} values for our full sample. 
We also employ composites of the VANDELS spectra to investigate the FUV spectral features as a function of \lyafesc, controlling for variations in \wlya. From these composites, we measure the equivalent width of low-ionization-state ISM absorption features ($W_{\rm{LIS}}$) to place constraints on {\lycfesc} using a low-redshift $W_{\rm{LIS}}-$\lycfesc \ calibration presented in \citet{saldana-lopez+22b}. 
Our main results can be summarised as follows:

\begin{enumerate}

    \item We find a positive correlation between {\lyafesc} and {\wlya} ($\simeq 10 \sigma$ significance), in which $f_{\rm{esc}}^{\rm{Ly\alpha}}$ monotonically increases from $f_{\rm{esc}}^{\rm{Ly\alpha}}\simeq0.04$ at $W_{\rm{\lambda}}(\rm{Ly\,\alpha})=10$ \AA\ to $f_{\rm{esc}}^{\rm{Ly\alpha}}\simeq0.1$ at $W_{\rm{\lambda}}(\rm{Ly\,\alpha})=25$ \AA.
    This represents the first measurement of the \lyafesc$-$\,\wlya\,relation at $z>4$ using \emph{individual} \lyafesc \ estimates.
    
    \item We show that the \lyafesc$-$\,\wlya\,relation does not evolve strongly from $z=0$ to $z=5$, and that the correlation holds down to \wlya$\,  \simeq 0$ \AA. Our results imply that the physical processes regulating the production and escape of \lya \ photons  from low metallicity,{\,} high\, $\xi_{\mathrm{ion}}$\, {\lya} emitters do not change significantly across $\simeq 90$ per cent of cosmic history.

    \item Using composite spectra, we show that as \lyafesc\, increases the strength of low-ionization-state ISM absorption lines decreases, consistent with a decrease in the covering fraction of neutral H{\sc i} gas.
    
    \item Using the relationship between the equivalent width of low-ionization absorption lines ($W_{\rm{LIS}}$) and \lycfesc \ derived from low redshift galaxies in the LzLCS survey \citep{saldana-lopez+22b}, we find that \lycfesc$\simeq 0.15^{+0.06}_{-0.04}$\lyafesc. Crucially, by constructing high- and low-\lyafesc composite spectra with the same \wlya, we demonstrate that the \lycfesc$-$\lyafesc\, relation is not being driven by a secondary correlation between \lyafesc\, and \wlya.
    
    \item We find that, at a given value of \wlya, the absolute \lycfesc \ values inferred from the low-redshift $W_{\rm{LIS}}-$\lycfesc \ calibration are a factor of $\geq 10$ lower than recent direct measurements of \lycfesc\, at $z>3$ \citep[e.g.,][]{steidel+18,pahl+21,begley+22}. This is similar to the offset reported in \citet{saldana-lopez+22b}.
    A number of systematic considerations may explain the discrepancy, but they remain to be fully understood. 
    We argue that caution must therefore be used in inferring \emph{absolute} values of \lycfesc \ from $W_{\rm{LIS}}$ measurements at high-redshift.
    
    In the future, JWST will offer improvements via direct spectroscopic and/or more accurate photometric \halpha\ measurements, which will lead to better constraints on {\lyafesc} for larger numbers of individual galaxies.
    
\end{enumerate}


\section*{Acknowledgements}
R. Begley, R. J. McLure, J. S. Dunlop, D.J. McLeod, C. Donnan and M.L. Hamadouche acknowledge the support of the Science and Technology Facilities Council. F. Cullen and T. M. Stanton acknowledge the support from a UKRI Frontier Re- search Guarantee Grant [grant reference EP/X021025/1]. A. C. Carnall thanks the Leverhulme Trust for their support via a Leverhulme Early Career Fellowship. This research made use of Astropy, a community-developed core Python package for Astronomy \citep{astropy13,astropy18},  NumPy \citep{numpy20} and SciPy \citep{scipy20}, Matplotlib \citep{matplotlib07}, IPython \citep{ipython07} and NASA’s Astrophysics Data System Bibliographic Services.

\section*{Data Availability}
The VANDELS survey is a European Southern Observatory Public Spectroscopic Survey. The full spectroscopic dataset, together with the complementary photometric information and derived quantities are available from \url{http://vandels.inaf.it}, as well as from the ESO archive \url{https://www.eso.org/qi/}.\\ For the purpose of open access, the author has applied a Creative Commons Attribution (CC BY) licence to any Author Accepted Manuscript version arising from this submission.



\bibliographystyle{mnras}
\bibliography{vandels_lya_ha} 



\appendix


\bsp	
\label{lastpage}
\end{document}